\documentclass[aps,prb,twocolumn,showpacs]{revtex4}

\usepackage{epsf}

\begin{document}

\title{Direct sampling of complex landscapes at low
temperatures:\\ the three-dimensional $\pm J$ Ising spin glass}

\author{Alexander K. Hartmann}
\email{hartmann@theorie.physik.uni-goettingen.de}
\affiliation{Institut f\"ur Theoretische Physik,University of
G\"ottingen, Bunsenstr. 9, 37073 G\"ottingen, Germany}

\author{Federico Ricci-Tersenghi}
\email{Federico.Ricci@roma1.infn.it}
\affiliation{Dipartimento di Fisica and INFM, Universit\`a di Roma
``La Sapienza'', Piazzale Aldo Moro 2, I-00185 Roma (Italy)}

\date{\today}

\begin{abstract}
A method is presented, which allows to sample directly low-temperature
configurations of glassy systems, like spin glasses.  The basic idea
is to generate ground states and low lying excited configurations
using a heuristic algorithm.  Then, with the help of microcanonical
Monte Carlo simulations, more configurations are found, clusters of
configurations are determined and entropies evaluated.  Finally
equilibrium configuration are randomly sampled with proper
Gibbs-Boltzmann weights.

The method is applied to three-dimensional Ising spin glasses with
$\pm J$ interactions and temperatures $T \le 0.5$.  The
low-temperature behavior of this model is characterized by evaluating
different overlap quantities, exhibiting a complex low-energy
landscape for $T>0$, while the $T=0$ behavior appears to be less
complex.
\end{abstract}


\maketitle

\section{Introduction}

Despite large efforts made by the scientists in the last two decades,
complex energy landscapes with many local minima and nested valleys,
like that of spin glasses~\cite{binder86}, still offer many relevant
questions to be answered.  These questions usually regard the lowest
energy levels of the landscape. The traditional numerical approach is
to apply a Monte Carlo (MC) simulation\cite{landau2000}.
Equilibration is tested by monitoring different average quantities as
a function of the number of MC steps. Equilibration can be assumed,
when the measured values of different runs, initially being far apart,
agree within error bars. Another approach\cite{KPY2001} is to
calculate one quantity, like the link overlap, in two different ways,
one time directly and one time depending on some other measured
quantity like the energy, and wait till both results agree.

Such a test is available only in special cases, e.g. for spin glasses
with a Gaussian distribution of the bonds.  Otherwise, one usually
waits till the quantity of interest does not show any more a time
dependence. Nevertheless, at low temperatures and with increasing
system size, equilibration becomes much harder and eventually, at very
low temperatures, is impossible.

In the very last years, a different approach has been proposed, namely
the calculation of ground-state (GS) and low-energy configurations.
Some characteristics of the low-energy landscape can be probed by the
application of suitable perturbations which slightly modify the
GS\cite{GroundStatePert}. But the full information on the
low-temperature behavior can be obtained only by an equilibrium
sampling of the system at a given temperature. Here we show, that by
calculating GS and excited states, one can directly sample very low
temperatures.  Several algorithms and heuristics \cite{opt-phys2001}
are available to obtain ground states and excited states. Some are
based again on Monte Carlo techniques like simulated annealing (SimA)
and parallel tempering (PT).  All these techniques have the drawback,
that it is impossible to obtained an unbiased, i.e. equilibrium sample
of configurations for $T\to 0$. For the MC methods, the reason is that
for larger systems and very low temperatures, equilibration times are
too long. We shall give below an example which shows for a $\pm J$
Ising spin glass, which exhibits an exponential ground state
degeneracy, that just obtaining ground states is much easier than
obtaining ground states with their proper statistics, i.e. each ground
state with the same probability. For other existing heuristics the
statistics of the configurations is influenced in an uncontrollable
way by the low-energy landscape.

In this work, a post-processing method is presented, which removes the
bias induced by the non-equilibrium low-temperature sampling and
allows to obtain a properly equilibrated state for systems having a
high degeneracy. The basic idea of the technique is to calculate
clusters of configurations, which are connected in configuration space
by zero-energy moves, e.g. zero-energy flips of spins in the Ising
spin-glass case. Next, the sizes of these clusters are estimated and
used to obtain an unbiased sample, where each cluster contributes with
a factor to the size of the cluster and to the Gibbs-Boltzmann (G-B)
weight.  This method has already been successfully applied to the
ground-state sampling of three-dimensional Ising $\pm J$ spin
glasses\cite{alex-equi}.  Here, the method is extended to the $T>0$
case and again applied to the $d=3$ $\pm J$ SG model.  Please note
that this approach works better and better with {\em decreasing}
temperature, hence is complementary to the MC technique, which suffers
from equilibration problems at low temperatures. But similar to MC,
one has to monitor some measured quantities as a function of some
parameters to establish equilibration, e.g. the number of clusters
found in the analysis as a function of the number of states included.
Also similar to MC, obtaining equilibrium becomes harder with
increasing system size. In this sense, the method is also not exact.
But in contrast to MC, ensuring equilibrium in this way is possible at
very low temperatures for larger systems (and becomes impossible for
higher temperatures), while for MC it is the other way round.

We apply the algorithm to three-dimensional Ising spin glasses.  The
EA model consists of $N=L^3$ Ising spins $s_i=\pm 1$ on a cubic
lattice with the Hamiltonian $H=-\sum_{\langle i,j \rangle} J_{ij} s_i
s_j$. The sum runs over all pairs of nearest neighbors $\langle i,j
\rangle$. The $J_{ij}$ are quenched random variables taking values
$J_{ij}=\pm 1$ with equal probability and satisfy the constraint
$\sum_{\langle i,j \rangle} J_{ij}=0$.  We apply periodic boundary
conditions in all directions.

In this work we show that the overlap distribution $P(q)$ at zero
temperature is qualitatively different from $P(q)$ at low but non-zero
temperature.  This means, even if there is an exponential number of GS
configurations, zero-temperature quantities may be very different from
those at any finite and small temperature.  In particular we will show
here that for the three-dimensional EA model, which has a finite
zero-temperature entropy, $P(q)$ is very narrow at exactly $T=0$,
while it is broad at any finite temperature.  We obtained the same
result for the box-overlap $P_{\rm box}(q)$. The picture resulting
from our findings is that of a large number of GS which are however
very close.  Nevertheless, quite different states can be easily found
once the first excited energy levels are considered.  This picture
agrees with the very recent MC results by Palassini and
Young~\cite{PY01}.

\begin{figure}[htb]
\begin{center}
\epsfxsize=0.75\columnwidth
\epsfbox{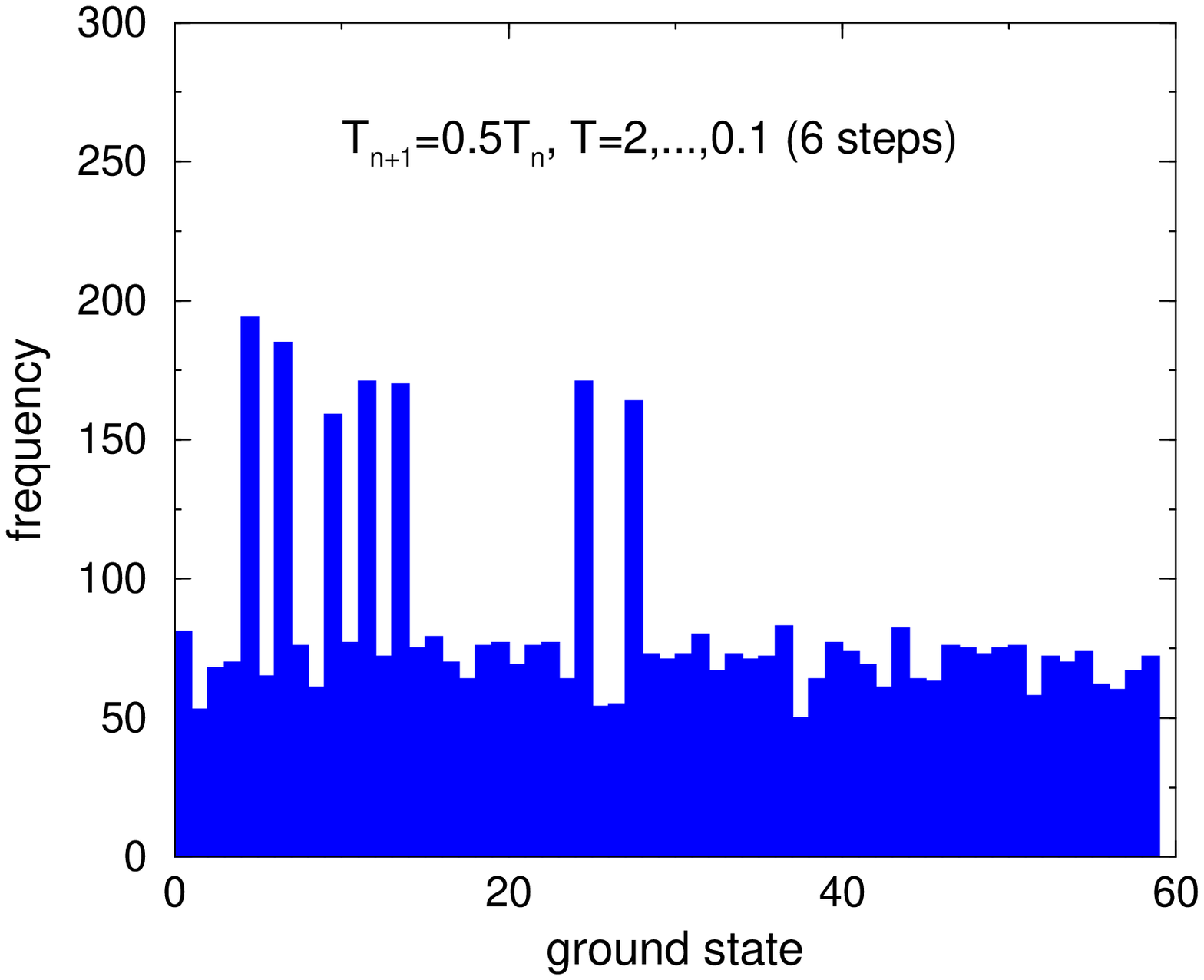}
\epsfxsize=0.75\columnwidth
\epsfbox{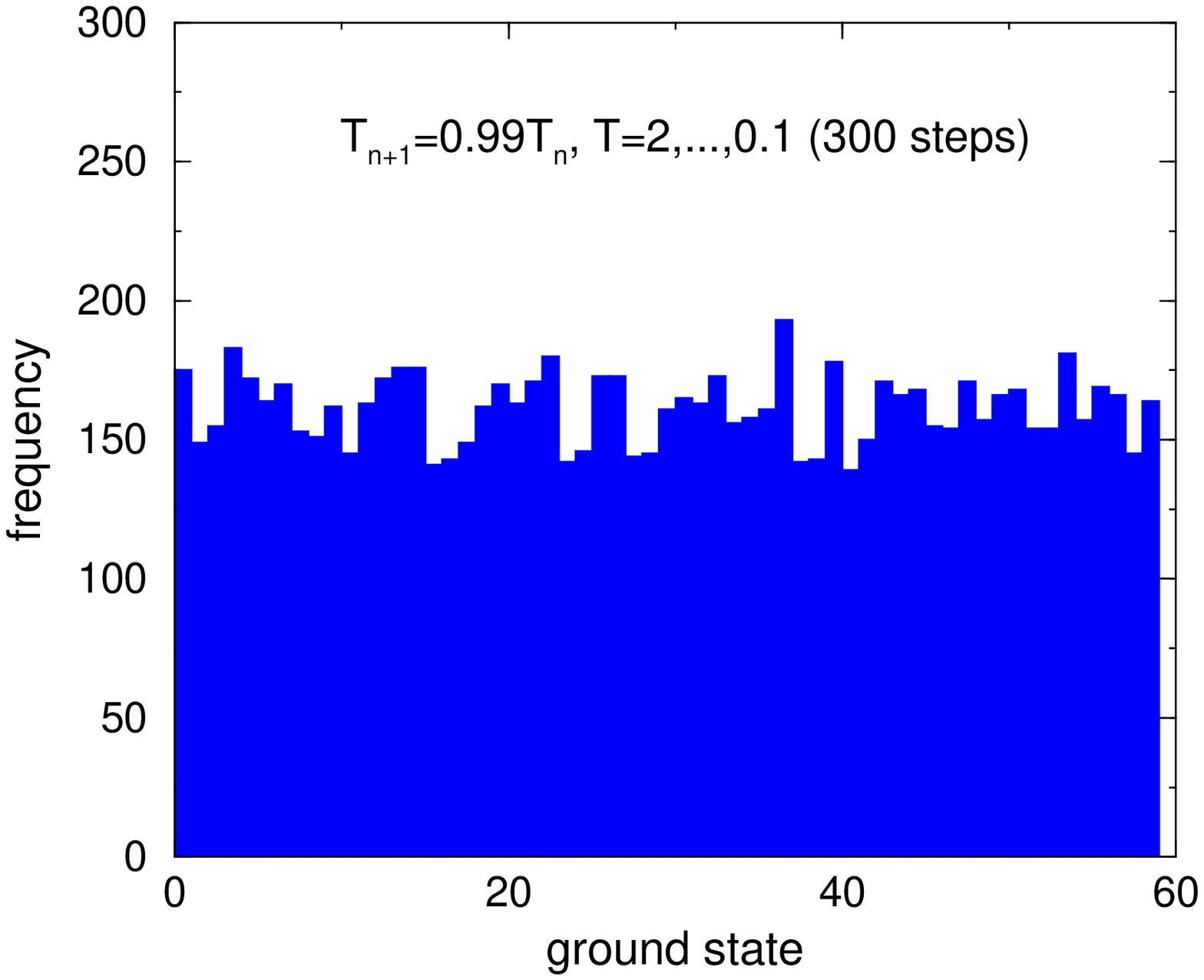}
\caption{Histogram of the number of times each GS is found with a SimA
simulation of $10^4$ independent runs for one $L=5$ realization of a
$\pm J$ Ising spin glass. The temperature was decreased according to
$T_{n+1}=bT_n$, with $T_0=2$ until $T=0.1$ is reached.  At each
temperature 10 MC sweeps were performed.  For the upper panel $b=0.5$,
while $b=0.99$ for the lower panel.}
\label{figSA}
\end{center}
\end{figure}

Before proceeding with our results and methods, we show, as a
motivation, results from applying the SimA method to one sample
realization of site $L=5$ of our model. We have performed $10^4$
independent runs of the SimA algorithm, starting with a temperature
$T_0=2$ and reducing the temperature according $T_{n+1}=bT_n$ until
$T=0.1$ is reached. Per temperature $10$ MC sweeps were performed. At
the end of the simulation, one randomly chosen configuration
exhibiting the lowest energy encountered during the run was
stored. After having performed $10^4$ runs, only the true ground
states were kept. A GS configuration and its mirror image, obtained by
reversing all states, are treated as being equivalent.  As it turns
out, the system has 59 distinct GS configurations. In Fig. \ref{figSA}
histograms of the number of times each GS has been found are displayed
for $b=0.5$ and $b=0.99$. One sees clearly that for $b=0.5$ different
GS configurations occur with different
frequencies\cite{footnote-cluster}, i.e. not all appear with the same
frequency as requested by the G-B distribution. When cooling much
slower, i.e. with $b=0.99$, all GS are almost equiprobable. This means
that just finding GS configurations is much easier than finding each
GS configuration with the correct probability.

For system sizes just slightly larger than $L=5$, the number of GS and
excited states is already huge (e.g. $\sim 10^{16}$ for $L=8$).  For
this system sizes it is impossible to obtain a histogram similar to
the one presented above. Consequently, it is impossible to determine
whether all GS are sampled with the correct statistics. This is even
more true for excited states. Please note that this is the same for
more elaborate algorithms like parallel
tempering\cite{juanjo2002}. Since, as already pointed out, at very low
temperatures and for system sizes like $L=10$ it is impossible to
equilibrate the system, other methods have to be applied. In this
paper, we present a post-processing tool, which allows to correct the
bias imposed by any algorithm and leads to an equilibrated sample. For
sizes up to $L=10$ and low temperatures up to $T<0.5$ the additional
effort is moderate, because only the few lowest levels of excited
states have to be considered.  For larger temperatures, the
post-processing methods becomes intractable, but then conventional MC
methods can be easily applied.

The rest of the paper is organized as follows. First, we explain the
algorithms we have applied. In the next section, we present the result
for the three-dimensional $\pm J$ spin glass. Finally, a summary and a
discussion are given.

\section{Algorithms}

The technique to obtained an equilibrated low-tempe\-rature sampling
consists of four steps:
\begin{enumerate}
\item Generate configurations for GS and the lowest levels of excitations.
\item On each energy level: group configurations into clusters. 
\item Calculate sizes of clusters.
\item Generate a sample of states for given temperature $T$, where
each cluster contributes with a weight proportional to its size and to
the G-B factor $\exp(-E/T)$, where $E$ is the energy of the
configurations in that cluster.
\end{enumerate}
Now all four steps are explained.

The basic method used here to generate the configurations is the
cluster-exact approximation (CEA) technique~\cite{alex2}, which is a
discrete optimization method~\cite{opt-phys2001} designed especially
for spin glasses.  In combination with a genetic
algorithm~\cite{pal96,michal92} this method is able to calculate true
GS~\cite{alex-stiff} up to $L=14$, as well as excited configurations
as a byproduct. Since the CEA technique is well established and
described in several sources, the details are skipped here.  For each
system and each energy level, we have generated 1000 configurations
with the pure genetic CEA algorithms. We will show below that this
number of configurations is sufficient up to $L=10$ and $T=0.5$.

By applying pure genetic CEA, one does not obtain the true
thermodynamic distribution~\cite{alex-false}, i.e.\ not all
configurations with the same energy contribute to physical quantities
proportional to the G-B weight. This means the genetic CEA algorithm
is biased.  For small system sizes up to $L=4$ it is possible to avoid
the problem by generating {\em all} low-energy configurations;
averages can be performed simply by considering each configuration
once, weighted with the G-B factor.  Since the degeneracy increases
exponentially with the number $N$ of spins and grows also strongly
with the energy level, a complete enumeration is not possible for
larger system sizes or higher energies.  Instead, one has to choose a
subset of all configurations, where each configurations contributes
with a probability proportional to the G-B weight.  The procedure
described here, consisting of steps 2-4 mentioned above, is applied to
ensure that all configurations appear with the correct probability in
this selection.  Please note that the following methods works for any
set of states, independently of the method which has been applied to
generate the states. I.e. also the results of many independent runs of
a low-temperature MC simulation can be treated, in case an
equilibration was not possible, e.g. for very low temperatures and
larger system sizes.

In step 2 of our method, we group the configurations into {\em
clusters} by performing the ballistic-search algorithm~\cite{alex-bs}:
All configurations which are accessible via flipping of spins having
zero local field (called {\em free spins} in the following), i.e.\
without changing the energy $E$, are considered to be in the same
cluster.  Please note that the Hamiltonian is symmetrical with respect
to flipping all spins simultaneously. Hence, for the rest of the paper
and for all analysis steps, a configuration and its mirror image are
regarded as being identical.  The final result is a list of different
clusters whose sizes are estimated as explained below.  This list does
not change if more than one configuration was initially found in the
same cluster, since these cases are recognized and correctly
handled. For completeness and to convince the reader that the method
indeed works, we present some details in the following.

The algorithm is applied independently for all configurations having
the same energy. The starting point is a set of $n_{\rm S}$
configurations.  For clarity, first the {\em straight-forward method}
to obtain the cluster structure is explained. This method will {\em
not} be applied finally. Afterward, the method actually used is
exposed.
 
The straight-forward construction starts with one arbitrary
configuration. It is the first member of the cluster. All
configurations which differ only by the orientation of one free spin
are called {\em neighbors}.  All the neighbors of the starting
configuration are added to the cluster. These neighbors are treated
recursively in the same way: All their neighbors which are yet not
included in the cluster are added, etc. After the construction of one
cluster is completed the construction of the next one starts with a
configuration, which has not been visited so far.

The construction of the clusters needs only linear computer-time as
function of $n_{\rm S}$ ($O(n_{\rm S})$), similar to the
Hoshen-Kopelman technique \cite{hoshen76}, because each configuration
is visited only once. Unfortunately the detection of all neighbors,
which has to be performed at the beginning, is of $O(n_{\rm S}^2)$
since all pairs of states have to be compared. Even worse, all
existing configurations of a given energy must have been calculated
before. As e.g. a $5^3$ system may exhibit already more than $10^5$ GS
and much more excited states, this algorithm is not suitable.

Instead we use the following technique, based on the {\em
ballistic-search} algorithm\cite{alex-bs}.  The basic idea of
ballistic search is to use a {\em test}, which tells whether two
configurations are in the same cluster. The test works as follows:
Given two independent replicas $\{\sigma_i^{\alpha}\}$ and
$\{\sigma_i^{\beta}\}$ let $D$ be the set of spins, which are
different in both states: $D\equiv \{i|\sigma_i^{\alpha}\neq
\sigma_i^{\beta}\}$. Now BS tries to build a path of successive flips
of free spins, which leads from $\{\sigma_i^{\alpha}\}$ to
$\{\sigma_i^{\beta}\}$ while using only spins from $D$.  In the
simplest version iteratively a free spin is selected randomly from
$D$, flipped and removed from $D$. This test does not guarantee to
find a path between two configurations which belong to the same
cluster, since it may depend on the order the spins are selected
whether a path is found or not. But, if a path is found, then it is
sure that both configurations belong to the same cluster. On the other
hand, if both configurations belong to the same cluster, then the
method finds a path with a certain probability which depends on the
size of $D$. It turns out that the probability decreases monotonically
with $|D|$.  For example for $N=8^3$ the method finds a path in 90\%
of all cases if the two states differ by 34 spins.  More analysis can
be found in \cite{alex-bs}.

The algorithm for the identification of clusters utilizes a collective
effect, to overcome the problem that sometimes a path is not found,
even if two configurations belong to the same cluster. It works as
follows: the basic idea is to let a configuration {\em represent} that
part of a cluster which can be found using BS with a high probability
by starting at this configuration. If a cluster is large it has to be
represented by a collection of states, such that the whole cluster is
``covered''. For example a typical cluster of a $8^3$ spin glass
consisting of $10^{16}$ ground states is usually represented by only
some few ground states (e.g. two or three).  A detailed analysis of
how many representing configurations are needed as a function of
cluster and system size can be found in \cite{alex-bs}. The details of
the algorithm are as follows: in memory a set of clusters consisting
each of a set of representing configurations is stored.  At the
beginning the cluster set is empty.  Iteratively all available
configurations $\{\sigma_i\}$ are treated: For all representing
configurations the BS algorithm tries to find a path to the current
configuration or to its inverse.  If no path is found, a new cluster
is created, which is represented by the actual configuration
treated. If $\{\sigma_i\}$ is found to be in exactly one cluster
nothing special happens. If $\{\sigma_i\}$ is found to be in more than
one cluster, it is called a {\em bridge configuration} and all these
clusters are merged into one single cluster, which is now represented
by the union of the states which have represented all clusters
affected by the merge.  After all configurations have been treated the
whole process is run again with the obtained set of clusters. This
allows to find bridge configurations which have not identified in the
first iteration, because accidentally only one cluster had been
created during the first iteration, at the time the configuration was
treated\cite{alex-bs}.

The BS identification algorithm has some advantages in comparison with
the straight-forward method: since each ground-state configuration
represents many ground states, the method does not need to compare all
pairs of states. Each state is compared only to a few number of
representing configurations. Thus, the computer time needed for the
calculation grows only a little bit faster than $O(n_{\rm S}n_C)$
\cite{alex-bs}, where $n_C$ is the number of clusters, which is much
smaller than $n_{\rm S}$.  Consequently, large sets of configurations,
which appear already for small system sizes like $N=5^3$, can be
treated. Furthermore, the cluster structure of even larger systems can
be analyzed, since it is sufficient to calculate a small number of
configurations per cluster.  The main point is that one has to be sure
that all clusters are identified correctly. This is not guaranteed
immediately, since for two configurations belonging to the same
cluster there is just a certain probability that a path of free
flipping spins connecting them is found. But this poses no problem,
because once at least one state of a cluster has been found, many more
states can be obtained easily by just performing a $E=$const
Monte-Carlo simulation starting with the initial state. Hence, one can
increasing the number of states available quickly. The probability
that all clusters have been identified correctly approaches very
quickly unity with increasing number of available states.  Detailed
tests can be found in \cite{alex-bs}. For all results presented here,
we have checked that the clusters do not change when doubling the
number of states.

\begin{figure}[htb]
\begin{center}
\epsfxsize=0.75\columnwidth
\epsfbox{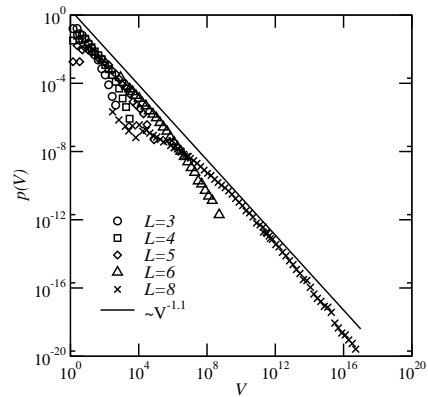}
\caption{Cluster-size distributions of GS clusters for small sizes
  $L=3$ to $L=8$. The straight line represents the function $2V^{-1.1}$.}
\label{figClusterDistribution}
\end{center}
\end{figure}

Furthermore, one has in principle to ensure that really all clusters
are found, which is simply done by calculating enough configurations,
but this is still only a tiny fraction of all configurations
\cite{alex-bs}. This time, the configurations must be obtained
independently, one cannot use the $E=$const MC simulation as above.
It is possible to obtain at least one configuration from each cluster
roughly up to size $L=8$ at GS level, resp. $L=6$ for first excited
states.  For sizes like $N=10^3$, the largest size we have treated in
this paper, the number of clusters is too large at any energy
level. But this is not a problem in principle because the
low-temperature behavior of these systems is dominated by large
clusters. As an example, in Fig. \ref{figClusterDistribution} the
probability {\em densities} of cluster sizes for GS clusters are
shown. The distributions are for small system sizes up to $L=8$, were
we can be fairly sure\cite{alex-sg234d} that all clusters have been
found\cite{footnote-cluster-size}. The distributions follow roughly an
algebraic decrease with a $p(V)\sim V^{-\alpha}$ behavior with
$\alpha\sim 1.1$. This dependence gets straighter with increasing
system size. We are interested in the contribution of a cluster of
{\em order} (or scale) of size $V$ to the behavior.  First, the
statistical weight of a cluster is proportional to the number of
states in the cluster, i.e. to the volume $V$. Second, each scale of
cluster sizes contributes proportional to the scale itself, because we
are integrating over all clusters of a given scale, i.e. this weight
is also proportional to $V$. (Or in other words, to translate the
probability densities into probabilities on a logarithmic scale, one
has to multiply with $V$.)  In total, clusters of sizes with scale $V$
contribute with weight $V^2p(V)=V^{2-\alpha}$. Since $\alpha\approx
1.1<2$, the largest scale clusters dominate the behavior.  On the
other hand, since $p(V)$ rapidly decreases, the {\em number} of these
dominating clusters is rather small, i.e. it is rather simple to
obtain an equilibrated sample of configurations. For the first excited
level we have found $\alpha=1.3<2$, while at higher excited levels the
number of clusters is too large to really find {\em all} of them. This
results indicates that at higher levels the distribution becomes
broader, which limits the application of the method to the lowest
level of excitations. This effect is studied below with more
detail. We have restricted our analysis to the first 4 levels of
excited states.

Please note that the CEA method generates configurations from larger
clusters with larger probability \cite{cea-analysis}, hence the large
and important clusters are encountered on average first in the
calculations. For the system sizes we have treated here, except $L=10$
and $T=0.5$, about 90\% of all contributing states are typically from
the top 5 largest clusters and further 5\% from the next 5 largest
clusters.  Then with the 1000 configurations we generated per energy
level, we encounter typically up to 100 clusters, and we can be pretty
sure that all thermodynamic relevant contributions are considered
within the level of accuracy given by our statistical fluctuations.
Only the results for $L=10$ and $T=0.5$, where higher level
excitations contribute significantly, may not be equilibrated. This is
demonstrated at the end of this section, after we have presented the
remaining parts of our algorithm.

The third step in the algorithm is the estimation of the cluster
sizes.  This works as follows.  Let $\cal C$ be a cluster we want to
measure in size and let's consider a random `reference configuration'
$\{r_i\}$ belonging to this cluster.  We define a test Hamiltonian
$\tilde{H}[s]=-\sum_i r_i s_i$ for $\{s_i\} \in \cal C$, being
$\tilde{E}(\beta)$ and $\tilde{S}(\beta)$ the average extensive energy
and entropy at inverse temperature $\beta$. Then the size of $\cal C$
is given by $\exp[\tilde{S}(0)]$.  Since the GS of this Hamiltonian is
unique (it is the reference configuration), i.e.\
$\tilde{S}(\infty)=0$, we obtain from the microcanonical definition of
the temperature $T=d\tilde{E}/d\tilde{S}$
\begin{eqnarray}
\tilde{S}(0) &=& \tilde{S}(0)-\tilde{S}(\infty) = \Delta\tilde{S} =
\int_{\tilde{E}(\infty)}^{\tilde{E}(0)} \beta\; d\tilde{E} = \nonumber
\\
&=& \int_0^\infty [\tilde{E}-\tilde{E}(\infty)]\; d\beta =
\int_0^\infty (\tilde{E}+N)\; d\beta \quad ,
\end{eqnarray}
where the previous last equality comes from an integration by parts
and the last equality from the substitution $\tilde{E}(\infty)=-N$.
In order to calculate this integral, we actually perform a fast MC
simulation restricted to configurations $\{s_i\} \in \cal C$ while
varying $w=\exp(-2\beta)$ in $[0,1]$ and measuring the average energy
$\tilde{E}$ as a function of $w$.  The final formula is the integral
of a smooth function $\Delta\tilde{S}= \int_0^1 \frac{N+\tilde{E}}{2w}
dw$. The number of MC sweeps applied per integration step was chosen
automatically by the program in a way that the resulting entropy did
not change by more than 5\% of the value when the number of MC sweeps
was doubled. I.e. the program started always with 10 MC sweeps,
calculated the entropy integral, then applied 20 MC sweeps and so
on. For small clusters, the calculation usually stopped after 20 MC
sweeps. For the largest clusters encountered here, the algorithm
stopped after the integration using 640 MC sweeps. We have also
checked, that for these cases the measured entropy did not depend
monotonically on the number of MC sweeps, i.e. we are sure that we did
not miss a systematic trend when stopping the calculation at one
point.

In principle, there could be high entropic barriers, which prevent the
size calculation from converging to the correct value. Fortunately,
the full algorithm is not susceptible to that problem. The reason is
that the BS clustering method uses single spin flips at constant
energy as well to determine the cluster structure, as described above.
This means, if two parts of a cluster are connected through a very
tiny path (the entropic barrier), which is not detected by the MC
integration, the clustering method is also not able to recognize both
subclusters as belonging to the same cluster. Hence, if both
subclusters are large, the genetic CEA method will have calculated
with high probability configurations from both subclusters. In the
analysis, because they are not identified as belonging to the same
cluster, they will appear as two independent large cluster, i.e.\ the
correct statistics is ensured at the end.  If on the other hand, one
subcluster is small, it has a negligible contribution to the overall
behavior, like other small clusters.

After estimating the cluster sizes, a certain number of configurations
is selected from each cluster, this is the last step of the algorithm
listed in the beginning of this section.  This number of
configurations is proportional to the size of the cluster and to the
G-B factor $\exp(-E/T)$.  It means that each cluster contributes with
its proper weight. This is possible for small temperatures and small
sizes, where only few low-energy levels contribute to the
thermodynamical behavior.

The selection of the configurations is done in a manner that many
small clusters may contribute as a collection as
well~\cite{alex-equi}.  For example, assume that 100 configurations
are selected from a cluster consisting of $10^{10}$ configurations,
then for a set of 500 clusters of size $10^7$ each (with the same
energy) a total number of 50 configurations is selected, i.e. 0.1
configurations per cluster on average.  The correct handling of such
situations is achieved by first sorting all clusters in ascending
order. Then the generation of configurations starts with the smallest
cluster. For each cluster the number of configurations generated is
proportional to its size, to $\exp(-E/T)$ and to a factor $f$. If the
number of configurations grows too large, only a certain fraction
$f_2$ of the configurations which have already been selected is kept,
the factor is recalculated ($f \leftarrow f*f_2$) and the process
continues with the next cluster.

The configurations representing the clusters are generated from the
initial configurations, obtained from the heuristic algorithm, by
microcanonical MC simulation, i.e.\ iteratively spins are randomly
selected and flipped if they are free.  Since within a cluster there
are no energy barriers, for the system sizes up to $L=10$, applying
100 MC sweeps ensures that all configurations within a cluster are
visited with the same frequency.

To summarize, by applying the algorithm presented here, each cluster
appears with a weight proportional to its size and to $\exp(-E/T)$ and
each configuration within a cluster appears with the same
probability. Therefore, on total, the correct thermodynamic
distribution is obtained.

\begin{figure}[htb]
\begin{center}
\epsfxsize=0.75\columnwidth
\epsfbox{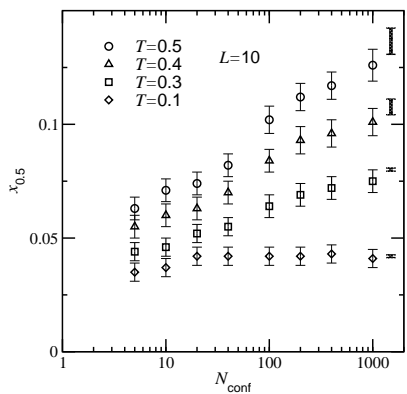}
\caption{Result for $x_{0.5}$ (see Eq. (\ref{eq:x05}) for definition)
as a function of the number $N_{\rm conf}$ of configurations included
per energy level in the analysis. The error bars at the right
represent the limiting values $N_{\rm conf}\to\infty$ obtained from
fitting the data points $N_{\rm conf}\ge 40$ to algebraic functions.
For small temperatures $T$, few configurations are sufficient while at
$T=0.5$ more than 1000 configurations are necessary.}
\label{figX05states}
\end{center}
\end{figure}

We have tested whether our generated data represents the equilibrium
behavior by calculating the small-overlap weight $x_{0.5}$, as defined
in the beginning of the next section in Eq. (\ref{eq:x05}).  $x_{0.5}$
is obtained for the largest system size $L=10$ and for different
temperatures $T$ as a function of the number of configurations $N_{\rm
conf}$ included in the analysis per energy level. The result is shown
in Fig.\ref{figX05states}.  Please note that the full analysis, as
explained in this section, has to be repeated independently for each
number $N_{\rm conf}$.  The configurations were taken in the order
they appeared in the generation using the genetic CEA, i.e. for a
small number of configurations, the large clusters are more likely to
be represented than the smaller clusters since genetic CEA
preferentially generates configurations from larger clusters.  One can
see that for low temperatures, even few generated configurations are
sufficient to yield the true behavior. Please note that the remaining
fluctuations are due to the fluctuations between the different samples
of configurations. The reason that few configurations are sufficient
here is that at low temperatures the GSs dominate and the number of GS
clusters is fairly small. With increasing temperature, excited states
become more important. For excited states, much more clusters
exists. Thus, more configurations must be included into the
analysis. This is visible in Fig. \ref{figX05states}, where at
e.g. $T=0.5$ $x_{0.5}$ depends strongly on $N_{\rm conf}$. For $N_{\rm
conf}=1000$ $T=0.5$ seems to be the borderline case, while for $T<0.5$
the result for $x_{0.5}$ seems to be converged (within error bars). We
have checked this explicitly by fitting algebraic functions to the
data points $N_{\rm conf}\ge 40$, resulting in an agreement within
error bars of the limiting value $N_{\rm conf}\to\infty$ with the
result we have obtained at $N_{\rm conf}\to\infty$.  Hence we can be
again confident that using 1000 configurations per energy level, the
results obtained here up to $L=10$ and $T<0.5$ represent the true
equilibrium behavior or, at least, is so close to the true result that
it cannot be distinguished from it at the level of accuracy determined
by the statistical fluctuations.  For smaller sizes, the number of
clusters is smaller on each energy level, which means that 1000
configurations per realization and energy level are sufficient for
even higher temperatures. But we restrict our analysis to $T \le 0.5$
here.

\begin{figure}[htb]
\begin{center}
\epsfxsize=0.75\columnwidth
\epsfbox{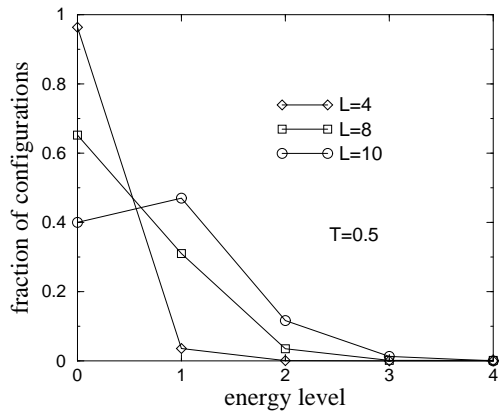}
\caption{Fraction of configurations sampled from each energy level at
$T=0.5$ for different system sizes. Energy level 0 is the ground
state.  Lines are guides to the eyes only.}
\label{figFractionL}
\end{center}
\end{figure}

Finally, in Fig.~\ref{figFractionL}, the fraction of configurations
sampled at $T=0.5$ for the different energy levels is shown for
different system sizes. For the smallest size $L=4$ almost only GS
configurations contribute to the thermodynamics, while increasing
system size higher energy configurations become more important. Please
note that only for $L=10$ configurations from excitation level 3
contribute. There the degeneracy is much larger than for the lower
levels. This explains, why the result for $L=10$ and $T=0.5$ is
probably not equilibrated.  The result of Fig.~\ref{figFractionL}
shows that, when studying the low-temperature behavior of glassy
systems, it is not sufficient to study just GS configurations since
the G-B factor and the size of the cluster (i.e.\ the entropy) must be
taken into account.  Nevertheless for low temperatures and not too
large system sizes, the energy levels which actually contribute to the
partition function are very few.

\section{Results}

We have calculated ground states and excited configurations up to
level four, for system sizes $L\le 10$. Up to 3000 realizations of the
disorder were considered (900 for the largest system size).  From the
set of configurations, samples of several hundred equilibrium
configurations were generated for temperatures $T\in[0,0.5]$.

\begin{figure}[htb]
\begin{center}
\epsfxsize=0.75\columnwidth
\epsfbox{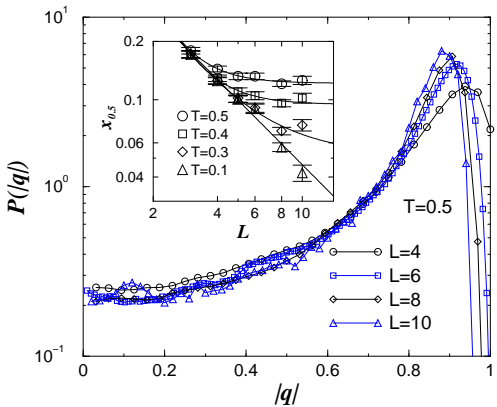}
\caption{Distribution $P(|q|)$ of overlaps at $T=0.5$ for different
system sizes. Lines are guides to the eyes only.  The inset shows the
average weight $x_{0.5}$ of the distribution for $|q|\le0.5$ as a
function of system size for $T=0.5, 0.4, 0.3, 0.1$. The lines
represent fits to functions of the form $x(L)=x^\infty+a\,L^\lambda$,
with $x^\infty \equiv 0$ and $\lambda=-1.10(5)$ for $T=0.1$,
$x^{\infty}= 0.051(13)$ for $T=0.3$, $x^{\infty}= 0.095(4)$ for
$T=0.4$ and $x^{\infty}= 0.122(4)$ for $T=0.5$.}
\label{figPq}
\end{center}
\end{figure}

For each disorder realization and each temperature, the distribution
$P_J(q)$ of overlaps $q \equiv \frac1N \sum_i s_i^{\alpha}
s_i^{\beta}$ was calculated, where $\{s_i^{\alpha}\}$,
$\{s_i^{\beta}\}$ are two different equilibrium configurations. In
Fig.~\ref{figPq} the disorder-averaged distribution
$P(|q|)=[P_J(|q|)]_{J}$ is shown for $T=0.5$, where $[\ldots]_J$
denotes the average over the quenched disorder. The long tail to $q=0$
seems to saturate at a finite weight, indicating the existence of a
complex low-energy landscape at finite temperatures. This can be seen
even better, by calculating the fraction
\begin{equation}
x_{q_0}=\int_{-q_0}^{q_0} P(q)\, dq
\label{eq:x05}
\end{equation} 
of overlaps smaller than $q_0$.  The result for $q_0=0.5$ is presented
in the inset of Fig.~\ref{figPq}.  For zero temperatures, where only
GS configurations are sampled, $x_{0.5}$ converges to 0 or to a very
small value~\cite{hed2001}.  The rate of convergence is described by
the finite-size dependence $x_{0.5}(L)\sim L^\lambda$. We find
$\lambda=-1.10(5)$, which is compatible with the predicted bound
$\lambda\le -1$ given by the ``TNT''-scenario\cite{krzakala2001}. In
Ref. \onlinecite{PY01} a larger value $\lambda=-0.90(10)$ was
found. This slight difference might be due to the different ensembles
studied, since in Ref.\onlinecite{PY01} the constraint $\sum_{\langle
i,j \rangle} J_{ij}=0$ was not applied.

Please note that for small temperatures we sample only GS
configurations, due to small system sizes.  For larger temperatures
$T\ge 0.3$, the asymptotic value of $x_{0.5}$ is clearly larger than
zero. Please note that the last point $L=10,T=0.5$ may not be
converged, as discussed above.  But, as you can see in
Fig.~\ref{figX05states}, the value of $x_{0.5}$ is an {\em increasing}
function of the number of states included in the calculation.  Hence,
the true result (we have obtained $x_{0.5}^{L=10}(0.5)=0.137(6)$ by
extrapolating $N_{\rm conf}\to \infty$ as
opposed to $0.126(7)$ found for $N_{\rm conf}=1000$) 
 is probably above our value, thus
supporting even more the conclusion that $x_{0.5}>0$.
 
Our results are quantitatively comparable to the data found in
Ref.~\onlinecite{PY01} which were obtained by a parallel-tempering MC
simulation.  Although the authors had no reliable criterion to check
equilibration of the system (in contrast to the case with Gaussian
distribution of the disorder~\cite{KPY2001}), by comparison with our
results it is very likely that in Ref.~\onlinecite{PY01} indeed
thermal equilibrium was obtained.

A non-trivial distribution of overlaps is not a sufficient criterion
for a complex energy landscape.  A qualitatively similar overlap
distribution with a nonzero weight for small values of $q$ would be
obtained also for a system, where various configurations differ by a
domain wall through the system at different positions, e.g.\ a
ferromagnet with antiperiodic boundary conditions in one
direction~\cite{fisher1987}.

\begin{figure}[htb]
\begin{center}
\epsfxsize=0.75\columnwidth
\epsfbox{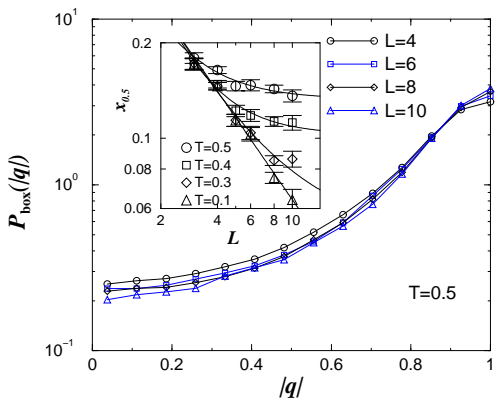}
\caption{Distribution $P_{\rm box}(|q|)$ of box overlaps at $T=0.5$
for different system sizes. Lines are guides to the eyes only.  The
inset shows the average weight $x_{0.5}$ of the distribution for
$|q|\le0.5$ as a function of system size for $T=0.5, 0.4, 0.3, 0.1$.
The lines represent fits to functions of the form
$x(L)=x^{\infty}_{\rm box}+a_b\,L^{\lambda_b}$, with $x^{\infty}_{\rm
box}\equiv 0$ and $\lambda_b=-0.86(5)$ for $T=0.1$, $x^{\infty}_{\rm
box}= 0.05(13)$ for $T=0.3$, $x^{\infty}_{\rm box}= 0.10(1)$ for
$T=0.4$ and $x^{\infty}_{\rm box}= 0.13(1)$ for $T=0.5$.}
\label{figPqBox}
\end{center}
\end{figure}

To rule out this scenario, we have calculated also the distributions
of box (or window) overlaps~\cite{newman1998,marinari1998}.  This
overlap is defined as usual, but restricted to a finite ``window'' of
volume $l\times l\times l$, with $l<L$ fixed independently of the
system size $L$.  Please note that for the aforementioned ferromagnet,
the distribution of box overlaps converges to a pair of delta
functions at $q=\pm 1$ when $L\to\infty$.  The result for $l=3$,
$T=0.5$ is exhibited in Fig.~\ref{figPqBox}.  At finite temperature,
similar to the conventional overlap, the low-$q$ tails seems to
saturate, but more slowly, at a non-zero weight with increasing
systems size.  This can be seen from the inset of Fig. \ref{figPqBox},
where $x_{0.5}$ is shown as a function of system size for
$T=0.1,0.3,0.4$ and $T=0.5$.  For $T \ge 0.3$, $x_{0.5}$ clearly
converges to a nonzero value.  Thus, we can conclude that indeed at
finite temperatures, three-dimensional spin glasses exhibit a complex
low-energy landscape.

Please note that the non-trivial behavior occurs for low temperatures,
probably for all temperatures $T>0$, which are sufficiently far away
from the phase transition $T_c\approx 1.1$.  Hence, the effects which
were found within a Migdal-Kadanoff approximation
scheme~\cite{kadanoff} are unlikely to explain the kind of behavior we
find.

\begin{figure}[htb]
\begin{center}
\epsfxsize=0.75\columnwidth
\epsfbox{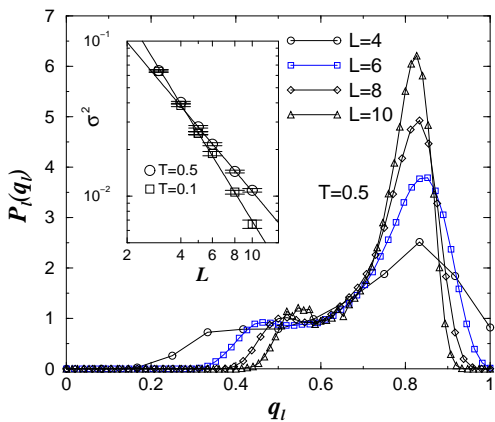}
\caption{Distribution $P_l(q_l)$ of link overlaps at $T=0.5$ for
different system sizes $L$. Lines are guides to the eyes only. The
inset shows the variance $\sigma^2$ as a function of system size for
$T=0.1, 0.5$. The lines represent fits to functions of the form
$\sigma^2(L)=a_l\,L^{\lambda_l}$ ($L>4$), with $\lambda_l=0.53(5)$ for
$T=0.1$ and $\lambda_l=0.27(1)$ for $T=0.5$.}
\label{figPqLink}
\end{center}
\end{figure}

Finally, we have computed the average distribution $P_l(q_l)$ of link
overlaps $q_l \equiv \sum_{\langle i,j \rangle} s^{\alpha}_i
s^{\alpha}_j s^{\beta}_i s^{\beta}_j$.  The result for $T=0.5$ and
different system sizes can be observed in Fig.~\ref{figPqLink}. The
distribution becomes narrower, but a second small peak seems to
emerge.  In the inset of Fig.~\ref{figPqLink} the finite-size
dependence of the variance $\sigma^2 = \int_{0}^1 (q-\bar{q})^2
P_l(q)\, dq$ is shown for different temperatures.  In all cases, the
width seems to converge toward zero.  Please note, however, that we
cannot exclude that the variance converges to a small but finite
value.  When we fit it to a function of the form $\sigma^2(L) =
\sigma^2_{\infty} + a_{\sigma}\, L^{\lambda_{\sigma}}$ we obtain, for
$T=0.5$, $\sigma^2_\infty = 0.0038(28)$ with $\chi^2$ per degree of
freedom of $0.1$, which is a very good fit.  Nevertheless, a
$P_l(q_l)$ consisting of two peaks at distance of 0.1 with weights 0.1
and 0.9 respectively has a variance $\sigma^2=0.0009$.

The behavior of $P_l(q_l)$ is quantitatively the same for
three-dimensional spin glasses with a Gaussian distribution of the
interactions~\cite{KPY2001}, which were found with a
parallel-tempering MC simulation.

\section{Summary}

Summarizing, we have presented an algorithm which allows to
investigate the low-temperature behavior of Ising systems with high
degeneracy by direct sampling of GS and excited configurations. The
basic idea is to generate configurations with any suitable algorithm,
group the configurations into clusters, measure the size of the
clusters and then obtain a very good estimate of the G-B measure, to
sample configurations with. Similar to MC, where one has to increase
the number of MC sweeps until the system is equilibrated, one has to
increase the number of independent configurations until the true
behavior is obtained. The main difference to MC techniques is that the
method presented here works better with decreasing temperature, while
MC equilibrates faster with increasing temperatures. In this sense
these methods are complementary.

We have applied the algorithm to study the low-temperature behavior of
three-dimensional $\pm J$ Ising spin glasses. We find that the
statistical properties of the exponentially many ground state
configurations are not representative of the low-temperature behavior.
In particular we have shown for the three-dimensional Edwards-Anderson
model that both the distributions of the overlap and of the
box-overlap seem to be very narrow functions at $T=0$, where only few
states contribute to the G-B measure, and broad for finite $T$.  Hence
the model does have a complex state space, which seems to become
trivial at $T=0$.  For this reason one is forced to probe the energy
landscape at $T>0$.  The distribution of the link-overlap seems to
develop a second peak, but the extrapolation of the asymptotic shape
is beyond our present computational capabilities.

\begin{acknowledgments}
The work was supported by the the {\em In\-ter\-diszi\-pli\-n\"a\-res
Zentrum f\"ur Wissenschaftliches Rechnen} in Heidelberg and the {\em
Paderborn Center for Parallel Computing} by the allocation of computer
time.  AKH acknowledges financial support from the DFG (Deutsche
Forschungsgemeinschaft) under grants Ha 3169/1-1 and Zi 209/6-1.
\end{acknowledgments}


\begin{thebibliography}{99}
\bibitem{footnote-cluster} A further analysis \cite{alex-bs} shows
that the GS of this system are organized in two clusters, where the GS
of each cluster are connected by zero-energy flips of single spins.
The GS found more often belong to one cluster, the other GS to the
second cluster.
\bibitem{footnote-cluster-size} On average, systems of size $L=8$ have
24.2 GS clusters. I.e. it is not difficult to be sure that all
clusters have been found.
\end{thebibliography}
\end{document}